\documentclass[preprintnumbers,amsmath,amssymb,floatfix,11pt,prd,onecolumn,showpacs,
superscriptaddress,nofootinbib]{revtex4}
\usepackage{graphicx}
\usepackage{epsfig}
\usepackage{bm}
\usepackage{amsfonts}

\begin{document}

\title{\large{\bf GRAVITATIONAL COLLAPSE WITH DARK ENERGY AND DARK MATTER IN HO$\check{\text R}$AVA-LIFSHITZ GRAVITY}}

\author{\textbf{Prabir Rudra}}\email{prudra.math@gmail.com}
\affiliation{Department of Mathematics, Pailan College of
Management and Technology, Bengal Pailan Park, Kolkata-700 104,
India.}
\author{\textbf{Ujjal Debnath}}\email{ujjaldebnath@gmail.com}
\affiliation{Department of Mathematics, Bengal Engineering and
Science University, Shibpur, Howrah-711 103, India.}

\begin{abstract}
In this work, the collapsing process of a spherically symmetric
star, made of dust cloud, is studied in Ho$\check{\text r}$ava
Lifshitz gravity in the background of Chaplygin gas dark energy.
Two different classes of Chaplygin gas, namely, New variable
modified Chaplygin gas and generalized cosmic Chaplygin gas are
considered for the collapse study. Graphs are drawn to
characterize the nature and to determine the possible outcome of
gravitational collapse. A comparative study is done between the
collapsing process in the two different dark energy models. It is
found that for open and closed universe, collapse proceeds with an
increase in black hole mass, the only constraint being that,
relatively smaller values of $\Lambda$ has to be considered in
comparison to $\lambda$. But in case of flat universe, possibility
of the star undergoing a collapse in highly unlikely. Moreover it
is seen that the most favourable environment for collapse is
achieved when a combination of dark energy and dark matter is
considered, both in the presence and absence of interaction.
Finally, it is to be seen that, contrary to our expectations, the
presence of dark energy does not really hinder the collapsing
process in case of Ho$\check{\text r}$ava-Lifshitz gravity.
\end{abstract}


\maketitle

\newpage

\section{INTRODUCTION}

\noindent

Cosmology has been a subject that has always attracted human mind
and as a result it has been an ever-developing field. Extensive
research has been carried out both theoretical and experimental
right from its early days. The most remarkable and significant
discovery in recent past in the field of cosmology is the
discovery of the fact, that our universe is undergoing an
accelerated expansion \cite{Perlmutter1, Spergel1}. This event has
shaken the traditional theories of cosmology right from their
roots. So Einstein's equation needed some serious revisions in
order to account for the observed cosmic acceleration.\\

As a result modified gravity theories came into existence. Loop
quantum gravity, brane-gravity, Ho$\check{\text r}$ava-Lifshitz
gravity \cite{Setare1, Jam1, Jam2, Karami1, Karami2}, etc. are
some of the modified gravity theories that was developed in the
recent past. In 2009, a new four dimensional gravity theory was
proposed by Ho$\check{\text r}$ava. The theory is devoid of full
diffeomorphism invariance but it has UV completeness. As a matter
of fact, the theory has three dimensional general covariance and
time re-parameterization invariance. In fact it is a
non-relativistic renormalizable quantum gravity theory possessing higher spatial derivatives.\\

Some cosmologists thought that the nature of the content of the
universe is more responsible than its geometry for the recent
cosmic acceleration. As a result the concept of dark energy (DE)
\cite{Riess1} was developed. It is basically a mysterious negative
pressure component which violates the strong energy condition i.e.
$\rho+3p<0$, thus causing accelerated expansion. Till date
numerous DE models have been proposed. Chaplygin gas model is one
such form of DE that has gained enormous popularity in the
cosmological society. Many Chaplygin gas models have appeared in
the scene, with extensive research. The earliest form was pure
Chaplygin gas (CG) \cite{Kamenshchik1, Gorini1}, which got
modified into generalized Chaplygin gas (GCG) \cite{Gorini2,
Alam1, Bento1, Barreiro1, Carturan1} and then into modified
Chaplygin gas (MCG) \cite{Benaoum1, Debnath1}. As time passed
variable Chaplygin gas (VMCG) \cite{Debnath2} and new variable
modified Chaplygin gas (NVMCG) \cite{Chakraborty1} came into
existence. In 2003, Gonz´alez-Diaz \cite{Gonzalez1} introduced the
generalized cosmic Chaplygin gas (GCCG) model \cite{Chakraborty2}.
The speciality of the model being that it can be made stable and
free from unphysical behaviours even when the vacuum fluid
satisfies the phantom energy condition. Another interesting
feature of this model is that it does not drive the universe
towards the Big-Rip singularity, unlike the other models
of Chaplygin gas.\\

The Equation of state (EoS) of NVMCG is given by
\cite{Chakraborty1},
\begin{equation}
p=A(a)\rho-\frac{B(a)}{\rho^{\alpha}}~,~~~~~~~~~~0\leq \alpha \leq
1
\end{equation}
Here we consider $A(a)=A_{0}a^{-n}$ and $B(a)=B_{0}a^{-m}$, where
$A_{0}$, $B_{0}$, $\alpha$, $m$ and $n$ are positive constants.
The motivation of the above choices has been discussed in ref
\cite{Chakraborty1}.

The Equation of state (EoS) of the GCCG model is
\cite{Gonzalez1,Chakraborty2}
\begin{equation}
p=-\rho^{-\alpha}\left[C+\left\{\rho^{(1+\alpha)}-C\right\}^{-\omega}\right]
\end{equation}
where $C=\frac{A}{1+\omega}-1$, with $A$ being a constant that can
take on both positive and negative values, and $-{l}<\omega<0$,
${l}$ being a positive definite constant, which can take on values
larger than unity. GCCG can explain the evolution of the universe
starting from the dust era to $\Lambda CDM$, radiation era, matter
dominated quintessence and lastly phantom era.

Recently cosmic coincidence problem and fine tuning problem have
crippled many dark energy models. In this regard interacting DE
models have been proposed \cite{Rud1, Rud2, Rud3, Rud4, Aslam1,
Pasqua1, Jamil1, Jamil2, Karami1}. The interacting models presents
a scenario of co-existence of DE and DM thus explaining the
present day universe in a far effective way.

Gravitational collapse is one of the most important problem in
classical general relativity for decades. The study of
gravitational collapse began with the pioneering work of
Oppenheimer and Snyder in 1939 \cite{Oppenhiemer1}, where they
studied the gravitational collapse of dust. From Penrose's Cosmic
censorship hypothesis (CCH) \cite{Penrose1}, we are made to
believe that any forms of gravitational collapse will resulting in
a singularity is destined to form Black holes. But the inquisitive
mind would like to know whether, and under what initial
conditions, gravitational collapse results in black hole (BH)
formation. Moreover, one would like to know if there are physical
collapse solutions that lead to naked singularities (NS), thus
violating the CCH. In last few years, there have been extensive
studies on gravitational collapse in order to investigate the
nature of the singularities \cite{Joshi,Ban,Debn,Dwiv}.

The collapsing process of a spherically symmetric star, made of
dust cloud, in the background of dark energy was studied for
Einstein's gravity \cite{Cai1,Cai2,Deb}, RSII Brane world model
\cite{Nath}, DGP Brane gravity and Loop Quantum gravity
\cite{Rudra1}. It was found that the presence of DE hinders the
collapsing procedure upto certain extent. Motivated by these
works, we intend to study the nature and outcome of gravitational
collapse of a star made up of DM in the background of DE (of
different forms), in Ho$\check{\text r}$ava-Lifshitz gravity
\cite{Hor1,Hor2,Hor3,Hor4,Hor5}.

The paper is organized as follows: We study the general
formulation of the collapsing process in section 2. Section 3 is
dedicated to the study of gravitational collapse in
Ho$\check{\text r}$ava-Lifshitz gravity. Collapse of dark matter
and dark energy with and without interactions are discussed. Two
types of dark energy i.e., new variable modified Chaplygin gas and
generalized cosmic Chaplygin gas are considered for study of
collapse. Section 4 deals with the detailed graphical analysis of
the plots generated. Finally the paper ends with a conclusion in
section 5.

%%%%%%%%%%%%%%%%%%%%%%%%%%%%%%%%%%%%%%%%%%%%%%%%%%%%%%%%%%%%%%%%%%%%%%%%%%%%%%%%%%%%%%%%%%%%%%%%%%%%%%%%%%%%%%%%%%%%%%%%%%%%%%%
\section{General Formulation of the Collapsing process}\label{chap02}
%%%%%%%%%%%%%%%%%%%%%%%%%%%%%%%%%%%%%%%%%%%%%%%%%%%%%%%%%%%%%%%%%%%%%%%%%%%%%%%%%%%%%%%%%%%%%%%%%%%%%%%%%%%%%%%%%%%%%%%%%%%
The flat, homogeneous and isotropic FRW model of the universe is
described by the line element
\begin{equation}\label{collapse3.3}
ds^{2}=dt^{2}+a^{2}(t)\left[dr^{2} +r^{2}\left(d\theta^{2}
+Sin^{2}\theta d\phi^{2}\right)\right]
\end{equation}
The energy conservation equation is given by
\begin{equation}\label{collapse3.4}
\dot{\rho}_{T}+ 3 \frac{\dot{a}}{a}(\rho_{T}+p_{T})=0
\end{equation}
with total density and pressure, $\rho_{T}=\rho_{M}+\rho_{E}$ and
$p_{T}=p_{M}+p_{E}$.

The interaction $Q(t)$ between DM and DE can be expressed as
\begin{equation}\label{collapse3.5}
\dot{\rho}_{M}+3 \frac{\dot{a}}{a} \rho_{M}=Q
\end{equation}
\begin{equation}\label{collapse3.6}
\dot{\rho}_{E}+ 3 \frac{\dot{a}}{a}(\rho_{E}+p_{E})=-Q
\end{equation}
Now, if we consider gravitational collapse of a spherical cloud
consists of above DM and DE distribution and is bounded by the
surface $\Sigma : r=r_{\Sigma}$ then the metric on it can be
written as
\begin{equation}\label{collapse3.7}
ds^{2}=dT^{2}-R^{2} (T) \{d\theta^{2} +Sin^{2} \theta d\phi^{2}\}
\end{equation}
Thus on $\Sigma : T=t$ and $R(T)=r_{\Sigma} a(T)$ where $
R(r,t)\equiv r a(t)$ is the geometrical radius of the two spheres
$t,r =$ constant. Also the total mass of the collapsing cloud is
given by \cite{Cai1,Cai2}
\begin{equation}\label{collapse3.8}
M(T) = \left. m(r,t)\right|_{r=r_{\Sigma}}=\left.\frac{1}{2} r^{3}
a \dot{a}^{2}\right|_{\Sigma}=\frac{1}{2}R(T)\dot{R}^{2}(T)
\end{equation}
The apparent horizon is defined as
\begin{equation}\label{collapse3.9}
R,_{\alpha} R,_{\beta} g ^{\alpha \beta}=0,~~{ i.e.,}~~
r^{2}\dot{a}^{2}=1
\end{equation}
So if $T=T_{AH}$ be the time when the whole cloud starts to be
trapped then
\begin{equation}\label{collapse3.10}
\left. \dot{R}^{2}(T_{AH})\right|_{\Sigma}
=r_{\Sigma}^{2}\dot{a}^{2}(T_{AH})=1
\end{equation}
As it is usually assumed that the collapsing process starts from
regular initial data so initially at $t=t_{i} ~(<T _{AH})$, the
cloud is not trapped i.e.,
\begin{equation}\label{collapse3.11}
r_{\Sigma}^{2} \dot{a}^{2}(t_{i})< 1,~~~~
 (r_{\Sigma}\dot{a}(t_{i})>-1)
\end{equation}
Thus if equation (\ref{collapse3.10}) has any real solution for
$T_{AH}$ satisfying (\ref{collapse3.11}) then black hole (BH) will
form, otherwise the collapsing process leads to a naked
singularity (NS). So the gravitational collapse and consequently
the formation of a BH solely depends upon the nature of root
obtained from equation (\ref{collapse3.10}). If any real solution
for $T_{AH}$ exists for equation (\ref{collapse3.10}) then
apparent horizon will be formed and thus a BH. If there is no real
solution the collapse is destined to result in a NS.

%%%%%%%%%%%%%%%%%%%%%%%%%%%%%%%%%%%%%%%%%%%%%%%%%%%%%%%%%%%%%%%%%%%%%%%%%%%%%%%%%%%%%%%%%%%%%%%%%%%%%%
\section{Gravitational Collapse in Ho$\check{\text R}$ava-Lifshitz gravity}\label{chap03}
%%%%%%%%%%%%%%%%%%%%%%%%%%%%%%%%%%%%%%%%%%%%%%%%%%%%%%%%%%%%%%%%%%%%%%%%%%%%%%%%%%%%%%%%%%%%%%%%%%%%%%%%
Here we briefly review the scenario where the cosmological
evolution is governed by Ho$\check{\text r}$ava-Lifshitz (HL)
gravity. The dynamical variables are the lapse and shift
functions, $N$ and $N_{i}$ respectively, and the spatial metric
$g_{ij}$. In terms of these fields the full metric is written as
\cite{Arn,Noj}
\begin{equation}
ds^{2}=-N^{2}dt^{2}+g_{ij}\left(dx^{i}+N^{i}dt\right)\left(dx^{j}+N^{j}dt\right)
\end{equation}
where the indices are raised and lowered using $g_{ij}$. The
scaling transformation of the coordinates reads: $t\rightarrow
l^{3}t$ and $x^{i}\rightarrow lx^{i}$.

The action of HL gravity is given by \cite{Hao}

$$I=dt\int dtd^{3}x\left(L_{0}+L_{1}+L_{m}\right)$$
$$L_{0}=\sqrt{g}N\left[\frac{2}{\kappa^{2}}\left(K_{ij}K^{ij}-\lambda
K^{2}\right)+\frac{\kappa^{2}\mu^{2}\left(\Lambda
R-3\Lambda^{2}\right)}{8\left(1-3\lambda\right)}\right]$$
\begin{equation}
L_{1}=\sqrt{g}N\left[\frac{\kappa^{2}\mu^{2}\left(1-4\lambda\right)}{32\left(1-3\lambda\right)}R^{2}-\frac{\kappa^{2}}{2\omega^{4}}\left(C_{ij}-\frac{\mu
\omega^{2}}{2}R_{ij}\right)\left(C^{ij}-\frac{\mu
\omega^{2}}{2}R^{ij}\right)\right]
\end{equation}

where $\kappa^{2}$, $\lambda$, $\mu$, $\omega$ and $\Lambda$ are
constant parameters, and $C_{ij}$ is Cotton tensor (conserved and
traceless, vanishing for conformally flat metrics). The first two
terms in $L_{0}$ are the kinetic terms, others in
$\left(L_{0}+L_{1}\right)$ give the potential of the theory in the
so-called ``detailed-balance" form, and $L_{m}$ stands for the
Lagrangian of other matter field. Comparing the action to that of
the general relativity, one can see that the speed of light and
the cosmological Newton's constant are

\begin{equation}
c=\frac{\kappa^{2}\mu}{4}\sqrt{\frac{\Lambda}{1-3\lambda}}~~,~~~~~~G_{c}=\frac{\kappa^{2}c}{16\pi
\left(3\lambda-1\right)}
\end{equation}
It should be noted that when $\lambda=1$, $L_{0}$ reduces to the
usual Lagrangian of Einstein's general relativity. Thus when
$\lambda=1$, the general relativity is approximately recovered at
large distances.

The field equations are

\begin{equation}
H^{2}+\frac{k}{a^{2}}=\frac{8\pi
G_{c}}{3}\left(\rho_{T}+\rho_{D}\right)
\end{equation}
and
\begin{equation}
\dot{H}+\frac{3}{2}H^{2}+\frac{k}{2a^{2}}=-4\pi
G_{c}\left(p_{T}+p_{D}\right)
\end{equation}
where
\begin{equation}
\rho_{D}=\frac{3\kappa^{2}\mu^{2}k^{2}}{8\left(3\lambda-1\right)a^{4}}+\frac{3\kappa^{2}\mu^{2}\Lambda^{2}}{8\left(3\lambda-1\right)}=\frac{1}{16\pi
G_{c}}\left(\frac{3k^{2}}{\Lambda a^{4}}+3\Lambda\right)
\end{equation}

and

\begin{equation}
p_{D}=\frac{\kappa^{2}\mu^{2}k^{2}}{8\left(3\lambda-1\right)a^{4}}-\frac{3\kappa^{2}\mu^{2}\Lambda^{2}}{8\left(3\lambda-1\right)}=\frac{1}{16\pi
G_{c}}\left(\frac{k^{2}}{\Lambda a^{4}}-3\Lambda\right)
\end{equation}

Using eqns. (12), (15) and (16) in (13) and (14), we get,

\begin{equation}
H^{2}+\frac{k}{a^{2}}=\frac{k^{2}c}{6\left(3\lambda-1\right)}\rho_{T}+\frac{1}{2}\left(\frac{k^{2}}{\Lambda
a^{4}}+\Lambda\right)
\end{equation}
and

\begin{equation}
\dot{H}+\frac{3}{2}H^{2}+\frac{k}{2a^{2}}=-\frac{k^{2}c}{4\left(3\lambda-1\right)}p_{T}-\frac{1}{4}\left(\frac{k^{2}}{\Lambda
a^{4}}-3\Lambda\right)
\end{equation}

In the following subsections, we shall discussed the collapse of
dark matter and dark energy in the form of new variable modified
Chaplygin gas and generalized cosmic Chaplygin gas separately and
also combination of dark matter and dark energy with and without
interactions.

%%%%%%%%%%%%%%%%%%%%%%%%%%%%%%%%%%%%%%%%%%%%%%%%%%%%%%%%%%%%%%%%%%%%%%%%%%%%%
\subsection{Collapse with Dark Matter}
%%%%%%%%%%%%%%%%%%%%%%%%%%%%%%%%%%%%%%%%%%%%%%%%%%%%%%%%%%%%%%%%%%%%%%%%%%%%%%%
Here $\rho_{M}\neq 0$, $p_{E}=\rho_{E}=0$. From the conservation
equation (3), we get
\begin{equation}
\rho_{M}=\frac{C_{0}}{a^{3}}
\end{equation}
Now using this relation in eqn.(17), we have,
\begin{equation}
\dot{a}=\sqrt{\frac{k^{2}cC_{0}}{6a\left(3\lambda-1\right)}+\frac{1}{2}\left(\frac{k^{2}}{\Lambda
a^{4}}+\Lambda\right)a^{2}-k}
\end{equation}
Therefore the expressions for the time gradient of the geometrical
radius of the collapsing cloud becomes,
\begin{equation}
\dot{R}(T)=r_{\Sigma}\sqrt{\frac{k^{2}cC_{0}}{6a\left(3\lambda-1\right)}+\frac{1}{2}\left(\frac{k^{2}}{\Lambda
a^{4}}+\Lambda\right)a^{2}-k}
\end{equation}
The mass of the collapsing cloud is given by
\begin{equation}
M(T)=\frac{1}{2}r_{\Sigma}^{3}\left[\frac{k^{2}cC_{0}}{6\left(3\lambda-1\right)}+\frac{1}{2}\left(\frac{k^{2}}{\Lambda
a^{4}}+\Lambda\right)a^{3}-ka\right]
\end{equation}

From the above solutions it is evident that as~~
$T\rightarrow\infty,~~a\rightarrow\infty,~~\rho_{M}\rightarrow0,~~\dot{R}(T)\rightarrow0,~~M(T)\rightarrow0$.
So we see that there is a tendency of matter density being
diminished as time passes, and finally it tends towards zero. The
time for formation of apparent horizon is given by the real root
of the equation, $\left. \dot{R}^{2}(T_{AH})\right|_{\Sigma}
=r_{\Sigma}^{2}\dot{a}^{2}(T_{AH})=1$ as given by equation
(\ref{collapse3.10}). Thus the corresponding expression for HL
gravity is,
\begin{equation}
r_{\Sigma}^{2}\left[\frac{k^{2}cC_{0}}{6a\left(3\lambda-1\right)}+\frac{1}{2}\left(\frac{k^{2}}{\Lambda
a^{4}}+\Lambda\right)a^{2}-k\right]=0
\end{equation}

\vspace{3mm}
\begin{figure}

\includegraphics[height=2in]{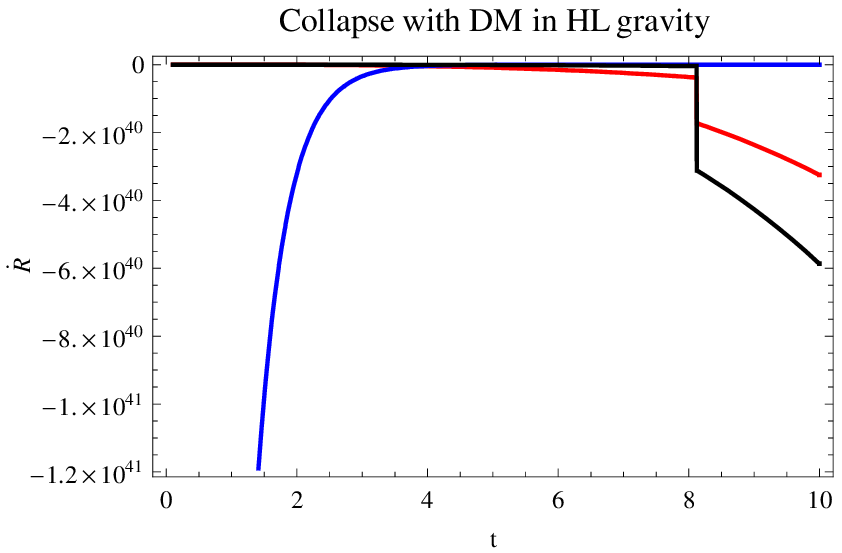}~~~~~~~~~~~~~~~~~~~~~\includegraphics[height=2in]{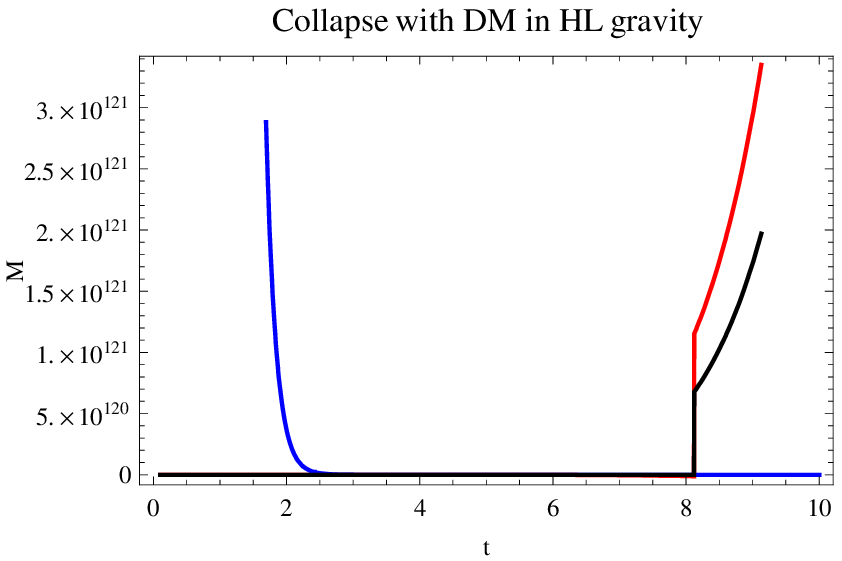}~~~\\
\vspace{1mm}
~~~~~~~~~~~~~Fig.1~~~~~~~~~~~~~~~~~~~~~~~~~~~~~~~~~~~~~~~~~~~~~~~~~~~~~~~~~~~~~~~~~~~~~~~~Fig.2~~~~~~~~~~~~~~~~~~~~\\

\vspace{3mm}

Fig 1 : The time derivative of the radius is plotted against time
for open universe (Red), flat universe (Blue) and closed universe
(Black) . The other parameters are considered as
$\Lambda=10,~ r=10,~ \lambda=100000,~ c=10,~ C_{0}=10$.\\
Fig 2 : The mass of the collapsing cloud is plotted against time
for open universe (Red), flat universe (Blue) and closed universe
(Black). The other parameters are considered as $\Lambda=10,~
r=10,~ \lambda=100000,~ c=10,~ C_{0}=10$.
\end{figure}

\vspace{1mm}

%%%%%%%%%%%%%%%%%%%%%%%%%%%%%%%%%%%%%%%%%%%%%%%%%%%%%%%%%%%%%%%%%%%%%%%%%%%%%%%%%%%%%%%%%
\subsection{Collapse with Dark Energy in the form of New Variable Modified Chaplygin Gas}
%%%%%%%%%%%%%%%%%%%%%%%%%%%%%%%%%%%%%%%%%%%%%%%%%%%%%%%%%%%%%%%%%%%%%%%%%%%%%%%%%%%%%%%%%
Here DE in the form of NVMCG is considered. So,
$\rho_{M}=0$,~~~~~$p=A(a)\rho-\frac{B(a)}{\rho^{\alpha}},~~~0\leq
\alpha \leq 1$ as given in equation (1). The solution for density
of NVMCG is obtained as \cite{Chakraborty1},

$$\rho_{nvmcg}=a^{-3}\exp\left(\frac{3A_{0}a^{-n}}{n}\right)\left[D_{0}+\frac{B_{0}}{A_{0}}\left(\frac{3A_{0}\left(1+\alpha\right)}{n}\right)^{\frac{3(1+\alpha)+n-m}{n}}\times
\Gamma \left(\frac{m-3\left(1+\alpha\right)}{n},\right.\right.$$
\begin{equation}\label{collapse3.24}
\left.\left.\frac{3A_{0}\left(1+\alpha\right)}{n}a^{-n}\right)\right]^{\frac{1}{1+\alpha}}
\end{equation}
where $D_{0}$ is the integration constant and $\Gamma (s,t)$ is
the upper incomplete gamma function.

The expressions for relevant physical quantities are

\begin{equation}\label{collapse3.21}
\dot{R}(T)=-r_{\Sigma}\sqrt{\frac{k^{2}ca^{2}\rho_{nvmcg}}{6\left(3\lambda-1\right)}+\frac{1}{2}a^{2}\left(\frac{k^{2}}{\Lambda
a^{4}}+\Lambda\right)-k}
\end{equation}
and
\begin{equation}\label{collapse3.22}
M(T)=\frac{1}{2}ar_{\Sigma}^{3}\left[\frac{k^{2}ca^{2}\rho_{nvmcg}}{6\left(3\lambda-1\right)}+\frac{1}{2}a^{2}\left(\frac{k^{2}}{\Lambda
a^{4}}+\Lambda\right)-k\right]
\end{equation}

The limiting values of the physical parameters are as follows:
$$\textbf{Case1}$$
$~~When~~a\rightarrow0:$ $~~~~\rho_{nvmcg}\rightarrow\infty$,
$~~~~\dot{R}\rightarrow-\infty$, $~~~~M(T)\rightarrow\infty$.
Obviously,~~~
$\frac{k^{2}ca^{2}\rho_{nvmcg}}{6\left(3\lambda-1\right)}+\frac{1}{2}a^{2}\left(\frac{k^{2}}{\Lambda
a^{4}}+\Lambda\right)\geq k_{max}=1.$

$$\textbf{Case2}$$
$~~When~~a\rightarrow\infty:$ $~~~~\rho_{nvmcg}\rightarrow 0$,
$~~~~\dot{R}\rightarrow-\infty$, $~~~~M(T)\rightarrow\infty.$

The cloud will start untrapped at the instant given by the real
roots of the following equation and gradually start to be trapped,
\begin{equation}
r_{\Sigma}^{2}\left[\frac{k^{2}ca^{2}\rho_{nvmcg}}{6\left(3\lambda-1\right)}+\frac{1}{2}a^{2}\left(\frac{k^{2}}{\Lambda
a^{4}}+\Lambda\right)-k\right]=1
\end{equation}

\vspace{3mm}

\begin{figure}

\includegraphics[height=2in]{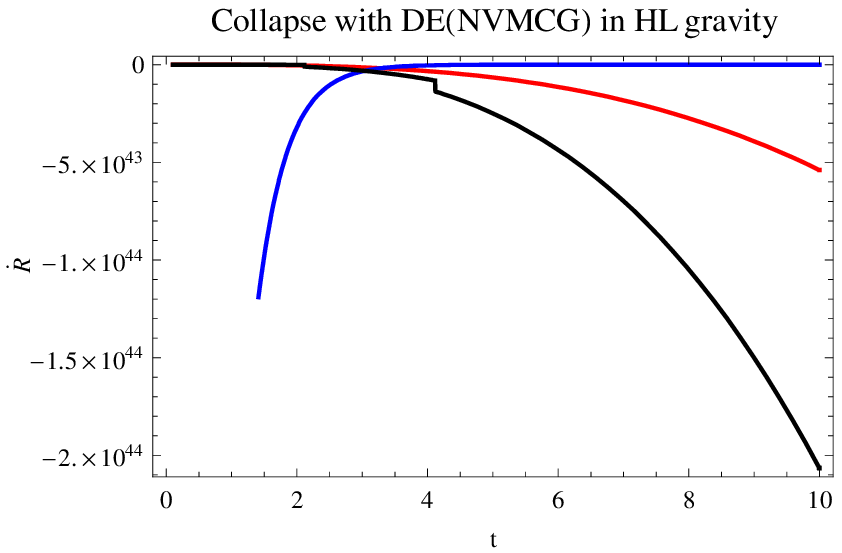}~~~~~~~~~~~~~~~~~~~~~\includegraphics[height=2in]{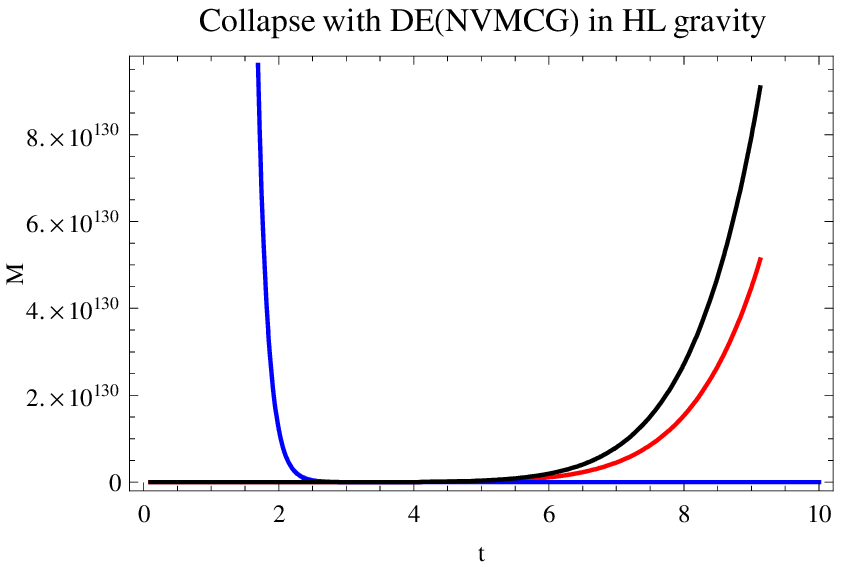}~~~\\
\vspace{1mm}
~~~~~~~~~~~~~Fig.3~~~~~~~~~~~~~~~~~~~~~~~~~~~~~~~~~~~~~~~~~~~~~~~~~~~~~~~~~~~~~~~~~~~~~~~~Fig.4~~~~~~~~~~~~~~~~~~~~\\

\vspace{3mm} Fig 3 : The time derivative of the radius is plotted
against time for open universe (Red), flat universe (Blue) and
closed universe (Black) . The other parameters are considered as
$\Lambda=1000,~ r=10,~ \lambda=1000000,~ c=10,~ A_{0}=1/3, B_{0}=3, D_{0}=10, \alpha=1/2, n=1, m=1 $.\\
Fig 4 : The mass of the collapsing cloud is plotted against time
for open universe (Red), flat universe (Blue) and closed universe
(Black). The other parameters are considered as $\Lambda=1000,~ r=10,~ \lambda=1000000,~ c=10,~ A_{0}=1/3, B_{0}=3, D_{0}=10, \alpha=1/2, n=1, m=1 $.\\

\end{figure}

\vspace{1mm}

%%%%%%%%%%%%%%%%%%%%%%%%%%%%%%%%%%%%%%%%%%%%%%%%%%%%%%%%%%%%%%%%%%%%%%%%%%%%%%%%%%%%%%%%%%%%%%%%%%%%%%%%%
\subsection{{\bf Effect of a combination of dark matter and New
variable modified Chaplygin gas}}
%%%%%%%%%%%%%%%%%%%%%%%%%%%%%%%%%%%%%%%%%%%%%%%%%%%%%%%%%%%%%%%%%%%%%%%%%%%%%%%%%%%%%%%%%%%%%%%%%%%%%%
\subsubsection{{\bf Case I : $Q = 0$ i.e., No Interaction Between
Dark Matter And Dark Energy:}}

In this case we consider the DE and DM to co-exist in a
non-interacting scenario. The energy density of DM is given by
eqn.(20) and the energy density of DE is given by eqn.(25).
Therefore in this case the total energy can be considered as the
sum total of the energy densities of DE and DM, i.e.
$\rho_{T}=\rho_{nvmcg}+\rho_{M}$. Using this in the field equation
for HL gravity, i.e. eqn.(17),we get the expressions of the
relevant parameters as given below.

The time gradient of the geometrical radius of the collapsing star
is given by,
\begin{equation}\label{collapse3.21}
\dot{R}(T)=-r_{\Sigma}\sqrt{\frac{k^{2}ca^{2}}{6\left(3\lambda-1\right)}\left(\rho_{nvmcg}+\frac{C_{0}}{a^{3}}\right)+\frac{1}{2}a^{2}\left(\frac{k^{2}}{\Lambda
a^{4}}+\Lambda\right)-k}
\end{equation}
and
\begin{equation}\label{collapse3.22}
M(T)=\frac{1}{2}ar_{\Sigma}^{3}\left[\frac{k^{2}ca^{2}}{6\left(3\lambda-1\right)}\left(\rho_{nvmcg}+\frac{C_{0}}{a^{3}}\right)+\frac{1}{2}a^{2}\left(\frac{k^{2}}{\Lambda
a^{4}}+\Lambda\right)-k\right]
\end{equation}
In this case the limiting values of the physical parameters are
given as follows,
$$\textbf{Case1}$$
$~~When~~a\rightarrow0:$ $~~~\rho_{nvmcg}\rightarrow\infty$,
$~~~~\dot{R}\rightarrow-\infty$, $~~~~M(T)\rightarrow\infty.$
Obviously,~~~
$\frac{k^{2}ca^{2}}{6\left(3\lambda-1\right)}\left(\rho_{nvmcg}+\frac{C_{0}}{a^{3}}\right)+\frac{1}{2}a^{2}\left(\frac{k^{2}}{\Lambda
a^{4}}+\Lambda\right)\geq k_{max}=1$

$$\textbf{Case2}$$
$~~When~~a\rightarrow\infty:$ $~~~~\rho_{nvmcg}\rightarrow 0$,
$~~~~\dot{R}\rightarrow-\infty$, $~~~~M(T)\rightarrow\infty.$

The cloud will start untrapped at the instant given by the real
roots of the following equation and gradually start to be trapped,
\begin{equation}
r_{\Sigma}^{2}\left[\frac{k^{2}ca^{2}}{6\left(3\lambda-1\right)}\left(\rho_{nvmcg}+\frac{C_{0}}{a^{3}}\right)+\frac{1}{2}a^{2}\left(\frac{k^{2}}{\Lambda
a^{4}}+\Lambda\right)-k\right]=1
\end{equation}

\vspace{3mm}
\begin{figure}

\includegraphics[height=2in]{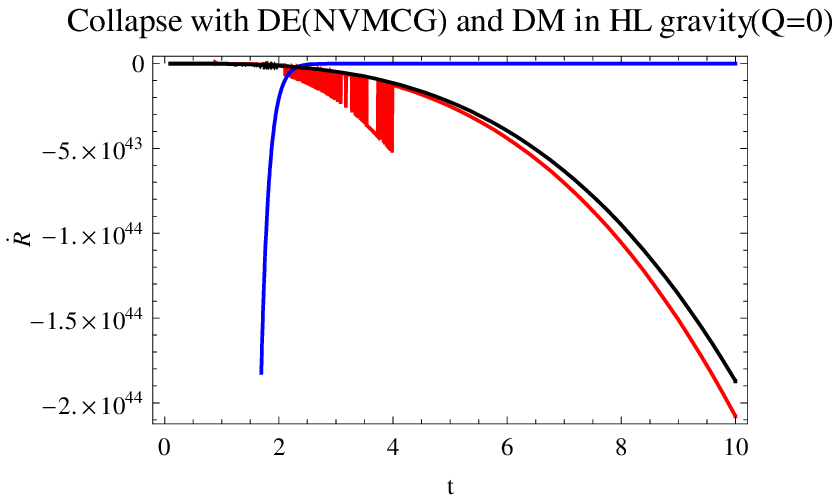}~~~~~~~~~~~~~~~~~~~~~\includegraphics[height=2in]{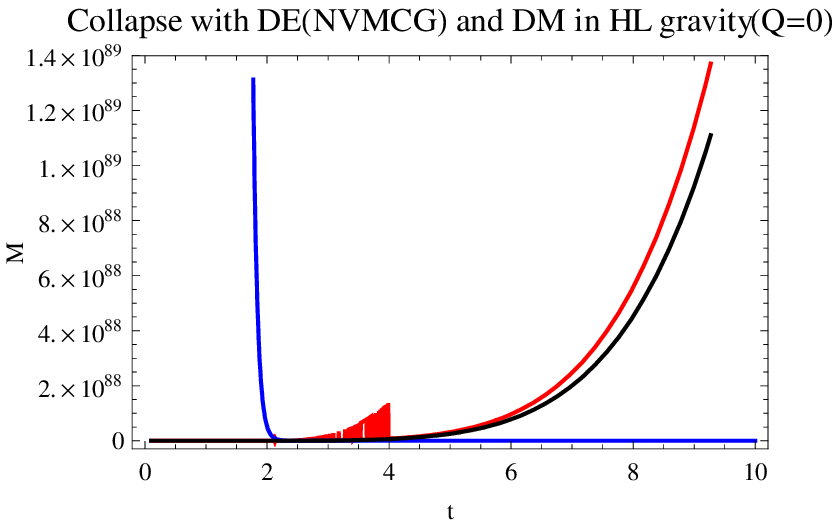}~~~\\
\vspace{1mm}
~~~~~Fig.5~~~~~~~~~~~~~~~~~~~~~~~~~~~~~~~~~~~~~~~~~~~~~~~~~~~~~~~~~~~Fig.6~~~~~~~~~~~~~~~~~~~~\\

\vspace{3mm} Fig 5 : The time derivative of the radius is plotted
against time for open universe (Red), flat universe (Blue) and
closed universe (Black) . The other parameters are considered as
$\Lambda=100,~ r=10,~ \lambda=100000,~ c=10,~ A_{0}=1/3, B_{0}=3, C_{0}=5, D_{0}=10, \alpha=1/2, n=1, m=1 $.\\
Fig 6 : The mass of the collapsing cloud is plotted against time
for open universe (Red), flat universe (Blue) and closed universe
(Black). The other parameters are considered as $\Lambda=100,~ r=10,~ \lambda=100000,~ c=10,~ A_{0}=1/3, B_{0}=3, C_{0}=5, D_{0}=10, \alpha=1/2, n=1, m=1 $.\\

\end{figure}

\vspace{1mm}

\subsubsection{{\bf Case II : $Q \neq 0$ i.e., Interaction Between
Dark Matter And Dark Energy:}}

Here we will use the assumption given by Cai and Wang in
\cite{Cai1, Cai2}. We consider
\begin{equation}
\frac{\rho_{nvmcg}}{\rho_{M}}=Ca^{3n'}
\end{equation}
where $C>0$ and $n'$ are arbitrary constants. We solve the
conservation equations (3) and (4) and get the following
expression for $\rho_{T}$ where $\rho_{T}=\rho_{nvmcg}+\rho_{M}$,

$$\rho_{T}=\exp\left( \frac{-a^{-n}}{n'\left(3n'-n\right)}\left\{a^{n}\left(n-3n'\right)\log \left(\frac{C+a^{-3n'}}{C\left(1+Ca^{3n'}\right)}\right)+3An'Ca^{3n'} ~_{2}F_{1}\left(1-\frac{n}{3n'},1,2-\frac{n}{3n'},-Ca^{3n'}\right)\right\}\right)$$
$$\left[C_{1}+\int_{1}^{t}-\frac{1}{1+Ca(u)^{3n'}}3C^{-\alpha}\exp\left( \frac{a(u)^{-n}}{n'\left(3n'-n\right)}\left\{a(u)^{n}\left(n-3n'\right)\log
\left(\frac{C+a(u)^{-3n'}}{C\left(1+Ca(u)^{3n'}\right)}\right)+\right.\right.\right.$$
$$\left.\left.\left. 3ACn'a(u)^{3n'}~_{2}F_{1}\left(1-\frac{n}{3n'},1,2-\frac{n}{3n'},-Ca^{3n'}\right)\right\}\right)\times a(u)^{-1-m-n-3\alpha
n'}\left(-Ba(u)^{n}\left(1+Ca(u)^{3n'}\right)^{\alpha}-\right.\right.$$
\begin{equation}
\left.\left.BCa(u)^{n+3n'}\left(1+Ca(u)^{3n'}\right)^{\alpha}\right)a'(u)du\right]
\end{equation}
where $_{2}F_{1}$ is the hypergeometric function and $C_{1}$ is
the integration constant. Using the relation
$\rho_{T}=\rho_{nvmcg}+\rho_{M}$. The expression for interaction
is given by,

$$Q=\frac{-3\rho_{T}}{\left(1+Ca^{3n'}\right)^{2}}\left[1+Ca^{3n'}+A_{0}Ca^{n+3n'}-B_{0}C^{-\alpha}a^{m-3n'\alpha}\left(\frac{1+Ca^{3n'}}{\rho_{T}}\right)^{\alpha+1}+Cn'a^{3n'}\right]\times$$
\begin{equation}
\sqrt{\frac{k^{2}c\rho_{T}}{6\left(3\lambda-1\right)}+\frac{1}{2}\left(\frac{k^{2}}{\Lambda
a^{4}}+\Lambda\right)-\frac{k}{a^{2}}}+3\sqrt{\frac{k^{2}c\rho_{T}}{6\left(3\lambda-1\right)}+\frac{1}{2}\left(\frac{k^{2}}{\Lambda
a^{4}}+\Lambda\right)-\frac{k}{a^{2}}}\left(\frac{\rho_{T}}{1+Ca^{3n'}}\right)
\end{equation}

Using the relation $\rho_{T}=\rho_{MCG}+\rho_{M}$ we get
\begin{equation}
\rho_{nvmcg}=\frac{Ca^{3n'}\rho_{T}}{1+Ca^{3n'}}~~,~~~~
~~~~~~~~~~~~~~~~~~\rho_{M}=\frac{\rho_{T}}{1+Ca^{3n'}}
\end{equation}
The gradient of scale factor is given by,
\begin{equation}
\dot{a}=-\sqrt{\frac{k^{2}ca^{2}}{6\left(3\lambda-1\right)}\rho_{T}+\frac{a^{2}}{2}\left(\frac{k^{2}}{\Lambda
a^{4}}+\Lambda\right)-k}
\end{equation}
where $\rho_{T}$ is given by eqn.(33).

The corresponding expressions for the time derivative of
geometrical radius is given by,
\begin{equation}
\dot{R}=-r\sqrt{\frac{k^{2}ca^{2}}{6\left(3\lambda-1\right)}\rho_{T}+\frac{a^{2}}{2}\left(\frac{k^{2}}{\Lambda
a^{4}}+\Lambda\right)-k}
\end{equation}
The expression for the mass is given by,
\begin{equation}
M(T)=\frac{1}{2}a^{2}r_{\Sigma}^{3}\left[\frac{k^{2}ca^{2}}{6\left(3\lambda-1\right)}\rho_{T}+\frac{a^{2}}{2}\left(\frac{k^{2}}{\Lambda
a^{4}}+\Lambda\right)-k\right]
\end{equation}
where $\rho_{T}$ is given by eqn.(33).

In this case the limiting values of the physical parameters are
given as follows,
$$\textbf{Case1}$$
$~~When~~a\rightarrow0:$ $~~~~\rho_{T}\rightarrow 0$,
$~~~~\dot{R}\rightarrow-\infty$, $~~~~M(T)\rightarrow
\frac{1}{4}r_{\Sigma}^{3}\frac{k^{2}}{\Lambda}.$ Obviously,~~~
$\frac{k^{2}ca^{2}}{6\left(3\lambda-1\right)}\left(\rho_{T}\right)+\frac{1}{2}a^{2}\left(\frac{k^{2}}{\Lambda
a^{4}}+\Lambda\right)\geq k_{max}=1.$

$$\textbf{Case2}$$
$~~When~~a\rightarrow\infty:$ $~~~~\rho_{nvmcg}\rightarrow 0$,
$~~~~\dot{R}\rightarrow-\infty$, $~~~~M(T)\rightarrow\infty.$

The cloud will start untrapped at the instant given by the real
roots of the following equation and gradually start to be trapped,
\begin{equation}
r_{\Sigma}^{2}\left[\frac{k^{2}ca^{2}}{6\left(3\lambda-1\right)}\left(\rho_{T}\right)+\frac{1}{2}a^{2}\left(\frac{k^{2}}{\Lambda
a^{4}}+\Lambda\right)-k\right]=1
\end{equation}

\vspace{3mm}
\begin{figure}

\includegraphics[height=2in]{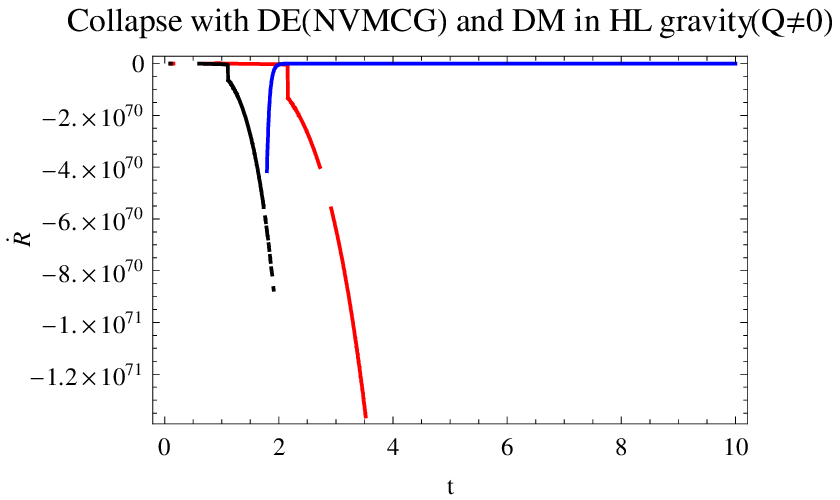}~~~~~~~~~~~~~~~~~~~~~\includegraphics[height=2in]{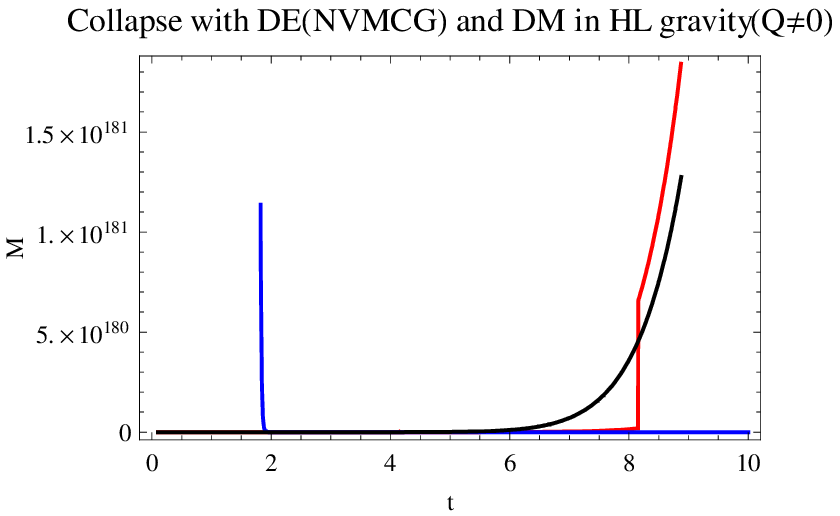}~~~\\
\vspace{1mm}
~~~~~~~Fig.7~~~~~~~~~~~~~~~~~~~~~~~~~~~~~~~~~~~~~~~~~~~~~~~~~~~~~~~~~~~~~~~~~~~~~~~~~~~~~~~~~Fig.8~~~~~~~~~~~~~~~~~~~~\\

\vspace{3mm} Fig 7 : The time derivative of the radius is plotted
against time for open universe (Red), flat universe (Blue) and
closed universe (Black) . The other parameters are considered as
$\Lambda=1000,~ r=10,~ \lambda=100000000,~ c=10, C=10, ~ A_{0}=1/3, B_{0}=3, \alpha=1/2, n=1, m=1, n'=2$.\\
Fig 8 : The mass of the collapsing cloud is plotted against time
for open universe (Red), flat universe (Blue) and closed universe
(Black). The other parameters are considered as $\Lambda=1000,~ r=10,~ \lambda=100000000,~ c=10,~ A_{0}=1/3, C=10, B_{0}=3, \alpha=1/2, n=1, m=1, n'=2 $.\\

\end{figure}

\vspace{1mm}

%%%%%%%%%%%%%%%%%%%%%%%%%%%%%%%%%%%%%%%%%%%%%%%%%%%%%%%%%%%%%%%%%%%%%%%%%%%%%%%%%%%%%%%%%
\subsection{Collapse with Dark Energy in the form of Generalized cosmic Chaplygin Gas}
%%%%%%%%%%%%%%%%%%%%%%%%%%%%%%%%%%%%%%%%%%%%%%%%%%%%%%%%%%%%%%%%%%%%%%%%%%%%%%%%%%%%%%%%%
Here DE in the form of GCCG is considered. So,
$\rho_{M}=0$,~~$p=-\rho_{GCCG}^{-\alpha}\left[C'+\left(\rho_{GCCG}^{1+\alpha}-C'\right)^{-\omega}\right]$
as given in equation (2) . The solution for density is given by
\cite{Chakraborty2}
\begin{equation}
\rho_{GCCG}=\left[C'+\left\{1+\frac{B'}{a^{3(1+\alpha)(1+\omega)}}\right\}^\frac{1}{1+\omega}\right]^\frac{1}{1+\alpha}~~,~~B'~
is~the~ integration~ constant.
\end{equation}
The expressions for the other physical quantities are given below,

\begin{equation}\label{collapse3.21}
\dot{R}(T)=-r_{\Sigma}\sqrt{\frac{k^{2}ca^{2}\left[C'+\left\{1+\frac{B'}{a^{3(1+\alpha)(1+\omega)}}\right\}^\frac{1}{1+\omega}\right]^\frac{1}{1+\alpha}}{6\left(3\lambda-1\right)}+\frac{1}{2}a^{2}\left(\frac{k^{2}}{\Lambda
a^{4}}+\Lambda\right)-k}
\end{equation}
and
\begin{equation}\label{collapse3.22}
M(T)=\frac{1}{2}ar_{\Sigma}^{3}\left[\frac{k^{2}ca^{2}\left[C'+\left\{1+\frac{B'}{a^{3(1+\alpha)(1+\omega)}}\right\}^\frac{1}{1+\omega}\right]^\frac{1}{1+\alpha}}{6\left(3\lambda-1\right)}+\frac{1}{2}a^{2}\left(\frac{k^{2}}{\Lambda
a^{4}}+\Lambda\right)-k\right]
\end{equation}

The limiting values of the physical parameters are as follows:
$$\textbf{Case1}$$
$~~When~~a\rightarrow0:$ $~~a\rightarrow
0:~~\rho_{GCCG}\rightarrow\infty,~~for ~1+\omega>0;~
\rho_{GCCG}\rightarrow\left(C'+1\right)^{\frac{1}{1+\alpha}},~
for~~1+\omega<0$,
$~~~~\dot{R}\rightarrow-\infty$, $~~~~M(T)\rightarrow\infty.$\\
Obviously,~~~
$\frac{k^{2}ca^{2}\left[C'+\left\{1+\frac{B'}{a^{3(1+\alpha)(1+\omega)}}\right\}^\frac{1}{1+\omega}
\right]^\frac{1}{1+\alpha}}{6\left(3\lambda-1\right)}+\frac{1}{2}a^{2}\left(\frac{k^{2}}{\Lambda
a^{4}}+\Lambda\right)\geq k_{max}=1.$

$$\textbf{Case2}$$
$~~When~~a\rightarrow\infty:$
$~~~~\rho_{GCCG}\rightarrow\infty,~~~~for
~1+\omega<0;~~~~\rho_{GCCG}\rightarrow\left(C'+1\right)^{\frac{1}{1+\alpha}},
for~~~~1+\omega>0$. $~~~~\dot{R}\rightarrow-\infty~~~~for
~~~1+\omega<0;~~~~\dot{R}\rightarrow
-r_{\Sigma}a\sqrt{\frac{k^{2}c\left(C'+1\right)^{\frac{1}{1+\alpha}}}{6\left(3\lambda-1\right)}
+\frac{\Lambda}{2}-k}~~~~for~~~~1+\omega>0$,
$~~~~M(T)\rightarrow\infty~~~~for
~~~1+\omega<0;~~~~M(T)\rightarrow
\frac{1}{2}a^{3}r_{\Sigma}^{3}\left[\frac{k^{2}c\left(C'+1\right)^{\frac{1}{1+\alpha}}}
{6\left(3\lambda-1\right)}+\frac{\Lambda}{2}-k\right]~~~~for~~~~1+\omega>0.$

The cloud will start untrapped at the instant given by the real
roots of the following equation and gradually start to be trapped,
\begin{equation}
r_{\Sigma}^{2}\left[\frac{k^{2}ca^{2}\left[C'+\left\{1+\frac{B'}{a^{3(1+\alpha)
(1+\omega)}}\right\}^\frac{1}{1+\omega}\right]^\frac{1}{1+\alpha}}{6\left(3\lambda-1\right)}+
\frac{1}{2}a^{2}\left(\frac{k^{2}}{\Lambda
a^{4}}+\Lambda\right)-k\right]=1
\end{equation}

\vspace{3mm}

\begin{figure}

\includegraphics[height=2in]{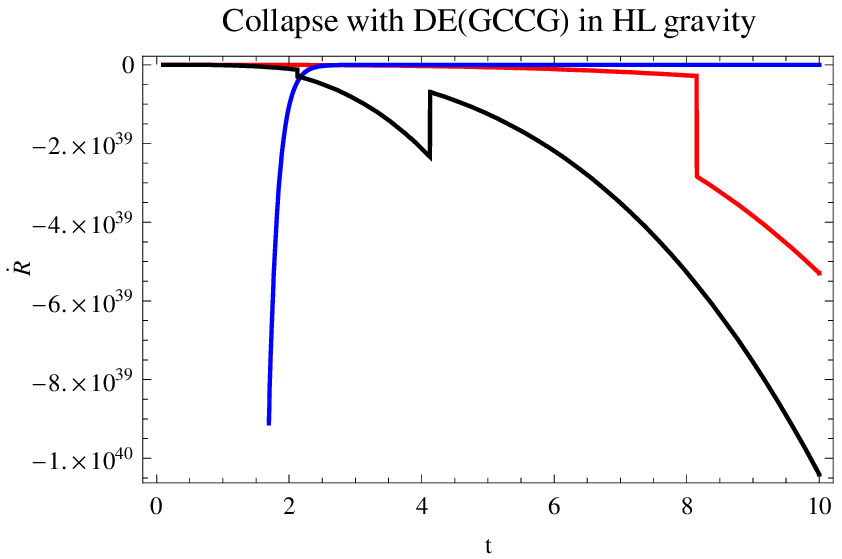}~~~~~~~~~~~~~~~~~~~~~\includegraphics[height=2in]{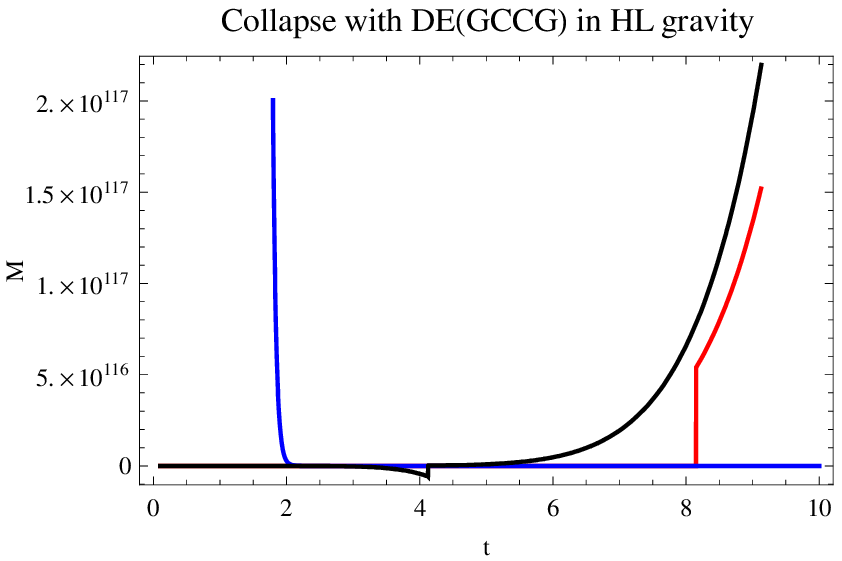}~~~\\
\vspace{1mm}
~~~~~~~Fig.9~~~~~~~~~~~~~~~~~~~~~~~~~~~~~~~~~~~~~~~~~~~~~~~~~~~~~~~~~~~~~~~~~~~~~~~~~~~~~~~~~Fig.10~~~~~~~~~~~~~~~~~~~~\\
Fig 9 : The time derivative of the radius is plotted against time
for open universe (Red), flat universe (Blue) and closed universe
(Black) . The other parameters are considered as
$\Lambda=100,~ r=10,~ \lambda=1000000, \omega=-2~, c=10, C'=5, B'=3, \alpha=1$.\\
Fig 10 : The mass of the collapsing cloud is plotted against time
for open universe (Red), flat universe (Blue) and closed universe
(Black). The other parameters are considered as $\Lambda=100,~ r=10,~ \lambda=1000000, \omega=-2~, c=10, C'=5, B'=3, \alpha=1$.\\
\vspace{3mm}

\end{figure}

\vspace{1mm}

%%%%%%%%%%%%%%%%%%%%%%%%%%%%%%%%%%%%%%%%%%%%%%%%%%%%%%%%%%%%%%%%%%%%%%%%%%%%%%%%%%%%%%%%%%%%%%%%%%%%%%%%%
\subsection{{\bf Effect of a combination of dark matter and
Generalized cosmic Chaplygin gas}}
%%%%%%%%%%%%%%%%%%%%%%%%%%%%%%%%%%%%%%%%%%%%%%%%%%%%%%%%%%%%%%%%%%%%%%%%%%%%%%%%%%%%%%%%%%%%%%%%%%%%%%
\subsubsection{{\bf Case I : $Q = 0$ i.e., No Interaction Between
Dark Matter And Dark Energy:}}

In this case we consider the DE and DM to co-exist in a
non-interacting scenario. The energy density of DM is given by
eqn.(20) and the energy density of DE is given by eqn.(40).
Therefore in this case the total energy can be considered as the
sum total of the energy densities of DE and DM, i.e.
$\rho_{T}=\rho_{GCCG}+\rho_{M}$. Using this in the field equation
for HL gravity, i.e. eqn.(17), we get the expressions of the
relevant parameters as given below.

The time gradient of the geometrical radius of the collapsing star
is given by,
\begin{equation}\label{collapse3.21}
\dot{R}(T)=-r_{\Sigma}\sqrt{\frac{k^{2}ca^{2}}{6\left(3\lambda-1\right)}\left(\left[C'+\left\{1+\frac{B'}{a^{3(1+\alpha)(1+\omega)}}\right\}^\frac{1}{1+\omega}\right]^\frac{1}{1+\alpha}+\frac{C_{0}}{a^{3}}\right)+\frac{1}{2}a^{2}\left(\frac{k^{2}}{\Lambda
a^{4}}+\Lambda\right)-k}
\end{equation}
and
\begin{equation}\label{collapse3.22}
M(T)=\frac{1}{2}ar_{\Sigma}^{3}\left[\frac{k^{2}ca^{2}}{6\left(3\lambda-1\right)}\left(\left[C'+\left\{1+\frac{B'}{a^{3(1+\alpha)(1+\omega)}}\right\}^\frac{1}{1+\omega}\right]^\frac{1}{1+\alpha}+\frac{C_{0}}{a^{3}}\right)+\frac{1}{2}a^{2}\left(\frac{k^{2}}{\Lambda
a^{4}}+\Lambda\right)-k\right]
\end{equation}
In this case the limiting values of the physical parameters are
given as follows,
$$\textbf{Case1}$$
$~~When~~a\rightarrow0:$ $~~~~\rho_{GCCG}\rightarrow\infty,~~~~for
~~~1+\omega>0;~~~~\rho_{GCCG}\rightarrow\left(C'+1\right)^{\frac{1}{1+\alpha}},~~~~for~~~~1+\omega<0$,
$~~~~\dot{R}\rightarrow-\infty$, $~~~~M(T)\rightarrow\infty.$\\
Obviously,~~~
$\frac{k^{2}ca^{2}}{6\left(3\lambda-1\right)}\left(\left[C'+\left\{1+\frac{B'}{a^{3(1+\alpha)(1+\omega)}}\right\}^\frac{1}{1+\omega}\right]^\frac{1}{1+\alpha}+\frac{C_{0}}{a^{3}}\right)+\frac{1}{2}a^{2}\left(\frac{k^{2}}{\Lambda
a^{4}}+\Lambda\right)\geq k_{max}=1.$

$$\textbf{Case2}$$
$~~When~~a\rightarrow\infty:$
$~~~~\rho_{GCCG}\rightarrow\infty,~~~~for
~~~1+\omega<0;~~~~\rho_{GCCG}\rightarrow\left(C'+1\right)^{\frac{1}{1+\alpha}},~~~~for~~~~1+\omega>0$,
$$~~~~\dot{R}\rightarrow-\infty~~~~for ~~~1+\omega<0;~~~~\dot{R}\rightarrow -r_{\Sigma}a\sqrt{\frac{k^{2}c\left(C'+1\right)^{\frac{1}{1+\alpha}}}{6\left(3\lambda-1\right)}+\frac{\Lambda}{2}-k}~~~~for~~~~1+\omega>0$$
$$~~~~M(T)\rightarrow\infty~~~~for ~~~1+\omega<0;~~~~M(T)\rightarrow \frac{1}{2}a^{3}r_{\Sigma}^{3}\left[\frac{k^{2}c\left(C'+1\right)^{\frac{1}{1+\alpha}}}{6\left(3\lambda-1\right)}+\frac{\Lambda}{2}-k\right]~~~~for~~~~1+\omega>0$$.
The cloud will start untrapped at the instant given by the real
roots of the following equation and gradually start to be trapped,
\begin{equation}
r_{\Sigma}^{2}\left[\frac{k^{2}ca^{2}}{6\left(3\lambda-1\right)}\left(\left[C'+\left\{1+\frac{B'}{a^{3(1+\alpha)(1+\omega)}}\right\}^\frac{1}{1+\omega}\right]^\frac{1}{1+\alpha}+\frac{C_{0}}{a^{3}}\right)+\frac{1}{2}a^{2}\left(\frac{k^{2}}{\Lambda
a^{4}}+\Lambda\right)-k\right]=1
\end{equation}

\vspace{3mm}
\begin{figure}

\includegraphics[height=2in]{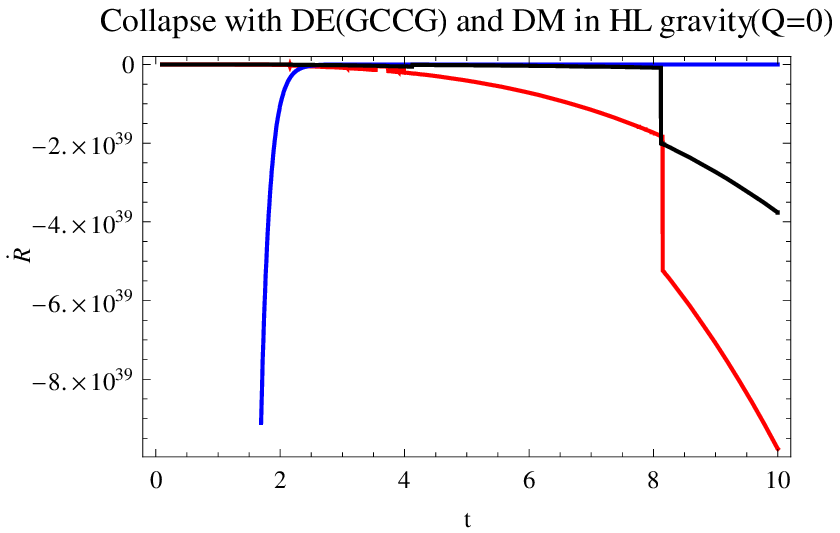}~~~~~~~~~~~~~~~~~~~~~\includegraphics[height=2in]{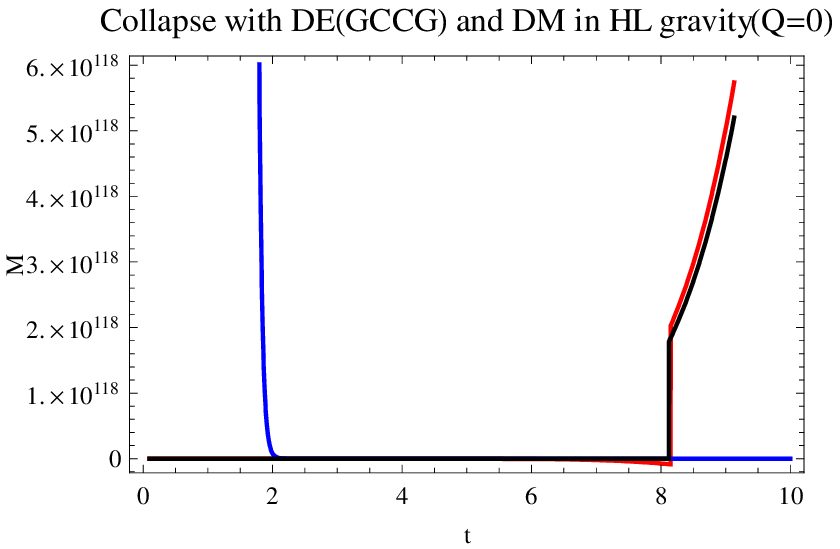}~~~\\
\vspace{1mm}
~~~~~~~Fig.11~~~~~~~~~~~~~~~~~~~~~~~~~~~~~~~~~~~~~~~~~~~~~~~~~~~~~~~~~~~~~~~~~~~~~~~~~~~~~~~~~Fig.12~~~~~~~~~~~~~~~~~~~~\\
\vspace{2mm}
Fig 11 : The time derivative of the radius is plotted
against time for open universe (Red), flat universe (Blue) and
closed universe (Black) . The other parameters are considered as
$\Lambda=100,~ r=10,~ \lambda=100000, \omega=-2~, c=10, C_{0}=10, C'=5, B'=3, \alpha=1/2$.\\
Fig 12 : The mass of the collapsing cloud is plotted against time
for open universe (Red), flat universe (Blue) and closed universe
(Black). The other parameters are considered as $\Lambda=100,~ r=10,~ \lambda=100000, \omega=-2~, c=10, C_{0}=10, C'=5, B'=3, \alpha=1/2$.\\
\vspace{3mm}

\end{figure}

\vspace{1mm}

\subsubsection{{\bf Case II : $Q \neq 0$ i.e., Interaction Between
Dark Matter And Dark Energy:}}

Just like the case of NVMCG, here also we consider
\begin{equation}
\frac{\rho_{GCCG}}{\rho_{M}}=C_{2}a^{3n''}
\end{equation}
where $C_{2}>0$ and $n''$ are arbitrary constants. Here
$\rho_{T}=\rho_{GCCG}+\rho_{M}$. Using this relation we get,
\begin{equation}
\rho_{GCCG}=\frac{C_{2}a^{3n''}\rho_{T}}{1+C_{2}a^{3n''}}~~,~~~~
~~~~~~~~~~~~~~~~~~\rho_{M}=\frac{\rho_{T}}{1+C_{2}a^{3n''}}
\end{equation}

Solving the conservation equations we get the expression for
$\rho_{T}$ as,
\begin{equation}
\rho_{T}=\frac{C_{3}\left(1+C_{2}a^{3n''}\right)^{\frac{1}{n''}}}{a^{3}}
\end{equation}
where $C_{3}$ is the integration constant. The expression for
interaction is given by,

$$Q=\frac{3\rho_{T}}{1+C_{2}a^{3n''}}\sqrt{\frac{k^{2}c\rho_{T}}{6\left(3\lambda-1\right)}+\frac{1}{2}\left(\frac{k^{2}}{\Lambda
a^{4}}+\Lambda\right)-\frac{k}{a^{2}}}\left[1-\frac{1}{1+C_{2}a^{3n''}}\left\{1+C_{2}a^{3n''}-\right.\right.$$
\begin{equation}
\left.\left.C_{2}a^{-3n''\alpha}\left(\frac{\rho_{T}}{1+C_{2}a^{3n''}}\right)^{-\alpha-1}\left(C'+\left(\left(\frac{C_{2}a^{3n''\rho_{T}}}{1+C_{2}a^{3n''}}\right)^{1+\alpha}-C'\right)^{-w}\right)+n''C_{2}a^{3n''}\right\}\right]
\end{equation}
where $\rho_{T}$ is given by eqn. (49). The expressions for the
other physical quantities are,
\begin{equation}
\dot{R}=-r\sqrt{\frac{k^{2}ca^{2}}{6\left(3\lambda-1\right)}\rho_{T}+\frac{a^{2}}{2}\left(\frac{k^{2}}{\Lambda
a^{4}}+\Lambda\right)-k}
\end{equation}
The expression for the mass is given by,
\begin{equation}
M(T)=\frac{1}{2}a^{2}r_{\Sigma}^{3}\left[\frac{k^{2}ca^{2}}{6\left(3\lambda-1\right)}\rho_{T}+\frac{a^{2}}{2}\left(\frac{k^{2}}{\Lambda
a^{4}}+\Lambda\right)-k\right]
\end{equation}
where $\rho_{T}$ is given by eqn.(49). In this case the limiting
values of the physical parameters are given as follows,
$$\textbf{Case1}$$
$~~When~~a\rightarrow0:$ $~~~~\rho_{T}\rightarrow \infty$,
$~~~~\dot{R}\rightarrow-\infty$, $~~~~M(T)\rightarrow \infty.$
Obviously,~~~
$\frac{k^{2}ca^{2}}{6\left(3\lambda-1\right)}\left(\rho_{T}\right)+\frac{1}{2}a^{2}\left(\frac{k^{2}}{\Lambda
a^{4}}+\Lambda\right)\geq k_{max}=1.$

$$\textbf{Case2}$$
$~~When~~a\rightarrow\infty:$ $~~~~\rho_{T}\rightarrow
C_{3}C_{2}^{\frac{1}{n''}}$,
$~~~~\dot{R}\rightarrow-r_{\Sigma}a\sqrt{\frac{k^{2}c}{6\left(3\lambda-1\right)}C_{3}C_{2}^{\frac{1}{n''}}+\frac{\Lambda}{2}}$,
$~~~~M(T)\rightarrow\frac{1}{2}a^{4}r_{\Sigma}^{3}\left[\frac{k^{2}c}{6\left(3\lambda-1\right)}C_{3}C_{2}^{\frac{1}{n''}}+\frac{\Lambda}{2}\right].$

The cloud will start untrapped at the instant given by the real
roots of the following equation and gradually start to be trapped,
\begin{equation}
r_{\Sigma}^{2}\left[\frac{k^{2}ca^{2}}{6\left(3\lambda-1\right)}\left(\rho_{T}\right)+\frac{1}{2}a^{2}\left(\frac{k^{2}}{\Lambda
a^{4}}+\Lambda\right)-k\right]=1
\end{equation}
where $\rho_{T}$ is given by eqn. (49).

\vspace{3mm}
\begin{figure}

\includegraphics[height=2in]{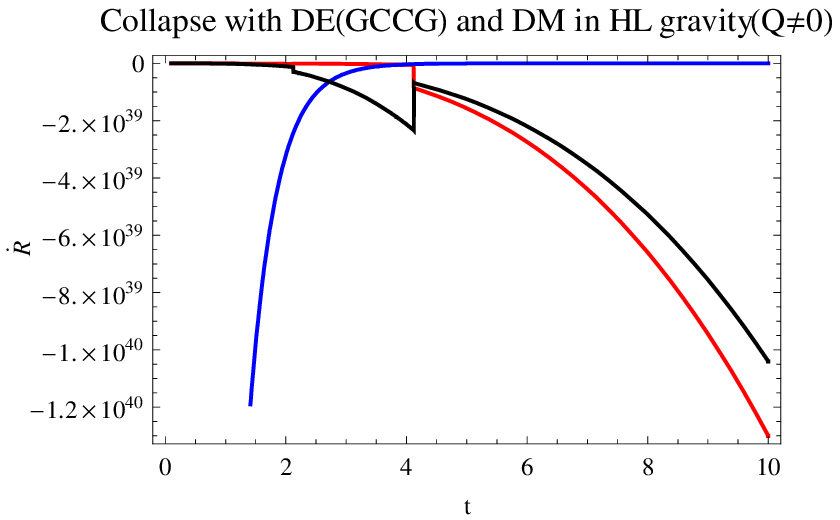}~~~~~~~~~~~~~~~~~~~~~\includegraphics[height=2in]{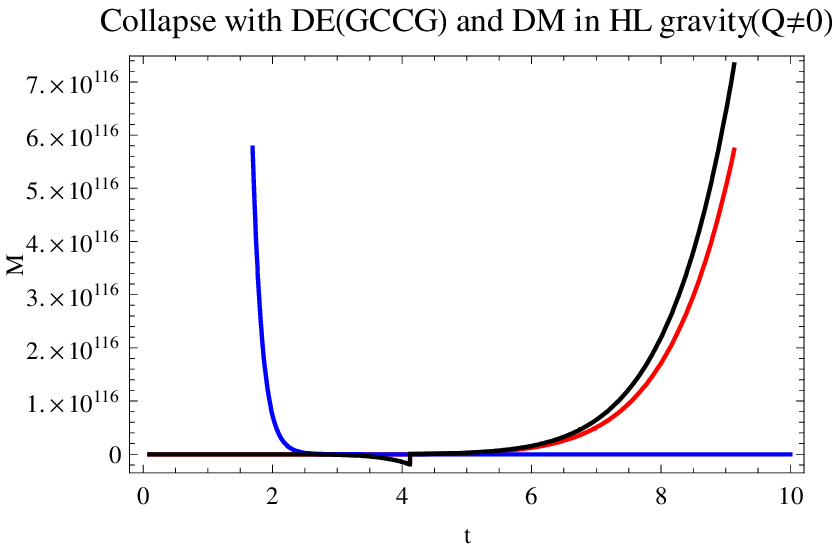}~~~\\
\vspace{1mm}
~~~~~~~Fig.13~~~~~~~~~~~~~~~~~~~~~~~~~~~~~~~~~~~~~~~~~~~~~~~~~~~~~~~~~~~~~~~~~~~~~~~~~~~~~~~~~Fig.14~~~~~~~~~~~~~~~~~~~~\\

\vspace{3mm} Fig 13 : The time derivative of the radius is plotted
against time for open universe (Red), flat universe (Blue) and
closed universe (Black) . The other parameters are considered as
$\Lambda=10,~ r=10,~ \lambda=100000000, C_{3}=2, C_{2}=5, c=10, n''=2, \alpha=1$.\\
Fig 14 : The mass of the collapsing cloud is plotted against time
for open universe (Red), flat universe (Blue) and closed universe
(Black). The other parameters are considered as $\Lambda=10,~ r=10,~
\lambda=100000000, C_{3}=2~, C_{2}=5, c=10, n''=2, \alpha=1$.\\
\end{figure}

\vspace{1mm}

\section{Graphical Analysis}
In this section we analyze the plots in detail. {\bf First of all
it should be mentioned that in all cases, the collapsing scenario
is realized for smaller values of $\Lambda$ and relatively much
higher values of $\lambda$.} In fig.1, a plot between the time
derivative of black hole radius, $\dot{R}$ against time is
provided for the collapsing scenario of dark matter. Trajectories
for open (Red), flat (Blue) and closed (Black) universe is shown
in a comparative scenario. It is seen that for flat universe
although the trajectory remains in the negative region yet its
increasing magnitude puts a lot of question in the feasibility of
the collapse. For open and closed universe, a perfectly collapsing
scenario is realized, with closed universe presenting the most
ideal conditions for gravitational collapse. In fig.2, black hole
mass, $M$ is plotted against time for an universe filled with dark
matter. Here also three different trajectories for open, flat and
closed universe are shown. We see that in case of open and closed
universe the mass increases, with the progression of collapse. The
mass is maximum in case of open universe. But in case of flat
universe the result is completely different. The collapse
progresses with a decrease in black hole mass. This is really
strange.\\

In fig. 3 and 4, respectively $\dot{R}$ and $M$ are plotted
against time for an universe filled with NVMCG type dark energy.
The results are quite similar to the previous case of dark matter
collapse, the only difference being that, the maximum mass is
obtained in case of closed universe, unlike the previous section.
{\bf If we compare the plots for NVMCG and DM it is evident that
introduction of DE does not affect the collapsing procedure in
case of closed universe, whereas in case of flat and open universe
some effect is visible.} In figs. 5 and 6, we plot $\dot{R}$ and
$M$ for the collapsing system when the universe is filled with a
combination of NVMCG and DM in a non-interacting scenario. Here we
see that collapse is favoured the most in case of open universe.
It is also seen that due to the co-existence of DE and DM, the
trajectories for open and closed universe almost coincide with
each other thus showing identical outcomes of collapse.
Interaction between DE and DM is considered and plots are
generated as shown in figs. 7 and 8. From fig.7, it is quite clear
that this is the most favourable environment for collapse to occur
as is evident from the steeply decreasing slopes of $\dot{R}$. It
is quite obvious that interaction between DE and DM is largely
responsible for this. The mass of the black holes increase quite
steeply in case of open and closed universe, but in case of flat
universe there is a decrease in mass just like the previous
sections.\\

In figs. 9 and 10, the plots are obtained for an universe filled
with GCCG as the DE. In this case as well tendency for collapse is
much less in case of flat universe compared to open and closed
universe. The plot for BH mass presents an identical scenario as
NVMCG. In figs. 11 and 12 the plots are generated for a
combination of GCCG and DM in a non-interacting scenario. Again
the striking feature is that the best collapsing scenario is
obtained for open universe unlike the previous case. Finally
considering interaction between GCCG and DM, figs. 13 and 14 are
obtained. If we closely look at the figures 5,6,7,8 and
11,12,13,14, {\bf we see that in case of GCCG the best collapsing
scenario is obtained in case of the combination of DE and DM in
the absence of interaction, much unlike NVMCG where the ideal
conditions were obtained in the presence of interaction.} This is
a really interesting result. Since the collapsing process for both
NVMCG and GCCG are identical, this difference can be phenomenally
attributed to the difference in their own internal mechanism as
DE.

\section{Conclusions}
Here we have studied the gravitational collapse of a spherically
symmetric dust cloud of finite radius, filled with homogeneous and
isotropic fluid. Two different types of DE fluids were considered
for our study, namely new variable modified Chaplygin gas and
generalized cosmic Chaplygin gas. The gravitational collapse of a
star filled with DM was studied in the background of DE, in
Ho$\check{\text r}$ava-Lifshitz gravity. The two different models
of DE was considered first independently and then in combination
with DM, with and without interaction. Both the studies are
compared and an attempt to obtain a meaningful result is made.
Relevant parameters (time derivative of radius and mass of the
collapsing cloud) were calculated and their variations with time
was plotted. A detailed graphical analysis was done to get a clear
understanding of the results obtained in the two different
models.\\

It was seen that for open and closed universe gravitational
collapse was accompanied with an increase of BH mass for both
NVMCG and GCCG. The above fact is quite understandable and
expected. But for flat universe occurrence of collapse is highly
unlikely. {\bf In case of open and closed universe, the only
constraint on collapse is that the values of $\lambda$ should be
relatively much higher than the values of $\Lambda$. For flat
universe, there is a bleak possibility of collapse irrespective of
any conditions.} The possible reasons are not very clear for the
time being. For a speculation we can attribute it to the quantum
effects of the Ho$\check{\text r}$ava-Lifshitz gravity. But
actually it remains an open question for the time being. The
hindrance in gravitational collapse due to the presence of DE is
not really prominent in our study, which was a major concern in
\cite{Rudra1}. {\bf Again it is a surprising feature of the
Ho$\check{\text r}$ava-Lifshitz gravity, that cosmic acceleration
is not really weakening the gravitational collapse, as it really
did in case of LQC and DGP brane in Rudra et al \cite{Rudra1}.}
This is really unexpected and the true reason remains an open
question for the time being.\\

{\bf Acknowledgement:}\\
The authors sincerely acknowledge the facilities provided by the
Inter-University Centre for Astronomy and Astrophysics (IUCAA),
Pune, India where a part of the work was carried out. Authors also
thank the anonymous referee for his/her invaluable comments that
helped to improve the quality of the paper.

\end{document}